# Coherent combining of self-cleaned multimode beams

Marc Fabert[1], Maria Săpânțan[1], Katarzyna Krupa[2], Alessandro Tonello[1], Yann Leventoux[1], Sébastien Février[1], Tigran Mansuryan[1], Alioune Niang[3], Benjamin Wetzel[1], Guy Millot[4,5], Stefan Wabnitz[6,7] and Vincent Couderc[1,*]

[1] Université de Limoges, XLIM, UMR CNRS 7252, 123 Avenue A. Thomas, 87060 Limoges, France
[2] Institute of Physical Chemistry, Polish Academy of Sciences, Warsaw, Poland
[3] Dipartimento di Ingegneria dell'Informazione, Università di Brescia, via Branze 38, 25123, Brescia, Italy
[4] Université de Bourgogne Franche-Comté, ICB, UMR CNRS 6303, 9 Avenue A. Savary, 21078 Dijon, France
[5] Institut Universitaire de France (IUF), 1 rue Descartes, 75005 Paris, France
[6] DIET, Sapienza University of Rome Via Eudossiana 18, 00184 Rome, Italy
[7] Novosibirsk State University, Pirogova 1, Novosibirsk 630090, Russia

e-mail* Vincent.couderc@xlim.fr

**A low intensity light beam emerges from a graded-index, highly multimode optical fibre with a speckled shape, while at higher intensity the Kerr nonlinearity may induce a spontaneous spatial self-cleaning of the beam [1,2]. Here, we reveal that we can generate two self-cleaned beams with a mutual coherence large enough to produce a clear stable fringe pattern at the output of a nonlinear interferometer. The two beams are pumped by the same input laser, yet are self-cleaned into independent multimode fibres. We thus prove that the self-cleaning mechanism preserves the beams' mutual coherence via a noise-free parametric process. While directly related to the initial pump coherence, the emergence of nonlinear spatial coherence is achieved without additional noise, even for self-cleaning obtained on different modes, and in spite of the fibre structural disorder originating from intrinsic imperfections or external perturbations. Our discovery may impact theoretical approaches on wave condensation [3-5], and open new opportunities for coherent beam combining [6-9].**

Interest in nonlinear multimode and multicore optical fibre systems has recently surged [10,11]. These fibres enable the new frontier of spatial-division-multiplexed optical communications [11-14], where nonlinearity remains the main limitation to system capacity [15]. Nonlinear multimode fibres (MMFs) permit orders of magnitude scaling of power from fibre lasers, the workhorse of industrial laser applications [16-18], and improved resolution for multiphoton microscopy and endoscopic imaging [11,19]. Interestingly, in multimode systems, the spontaneous emergence of beam cleanup has been observed, potentially mediated by photon interactions with optical [20] or acoustic [21] phonons, or by direct photon-photon interactions [1,2,22-24]. In the latter case, the collective beam dynamics has a direct hydrodynamic analogy with the order parameter of a three-dimensional boson gas: multimode beam condensation is predicted to emerge from turbulent wave mixing [3-5].

While photons do not interact in vacuum, the interplay between photons of a light beam is however possible in a nonlinear medium. In this regime, the hydrodynamic features of light are well-known in guided wave optics: in the framework of nonlinear interactions within multidimensional systems such as MMFs, optical beams are often referred to as a *"fluids of light"* [25]. Our findings confirm the hydrodynamic picture of multimode beam propagation as a fluid of light. Indeed, Maxwell's equations in the paraxial approximation lead to the Gross-Pitaievskii equation [26], which describes the mean-field dynamics of a three-dimensional Bose-Einstein condensate density (i.e., order parameter): the linear refractive index profile plays the role of a trapping potential, and cubic nonlinearity corresponds to boson-boson interactions [25]. In contrast with photorefractive crystals [27], a local and instantaneous nonlinearity is involved in optical fibres: the propagation of the photon fluid occurs in a three-dimensional space, where physical time $t$ plays the role of a third coordinate. This entails a full three-dimensional thermalization of the photon gas: it has been shown that cooling in transverse dimensions is accompanied by temperature increase in the temporal dimension [28-29].

In this framework, self-cleaned multimode coherent light sources open a new perspective for noise-immune coherent beam combining systems [6-9], which are at the cornerstone of high-power fibre laser technologies for a variety of applications, ranging from laser fusion to lidar-based environmental monitoring. From a fundamental physics viewpoint, our results indicate the possibility of maintaining the quantum coherence of a fluid of light over macroscopic evolution times. Nonlinear multimode systems could thus constitute an ideal test-bed to investigate many-body phenomena in quantum statistical mechanics, such as strong quantum fluctuations of the photon-field phase, and Hawking radiation in analog black-hole configurations [25].

In experiments, we coupled a Gaussian optical beam with central wavelength at 1064 nm and pulse duration of 60 ps into a multimode Mach-Zehnder interferometer followed by a dispersive element, implemented by means of a bulk grating to perform a 1D/1D spatial/spectral analysis of the multimode output beam (see Fig.1a). Specifically, the initial laser beam was launched into two 12 m-long spans of graded-index (GRIN) MMFs, respectively placed in each arm of the interferometer, and their output was spatially superimposed with a small angle (~ 4°) after propagation through the two independent GRIN MMFs sections (see Methods).

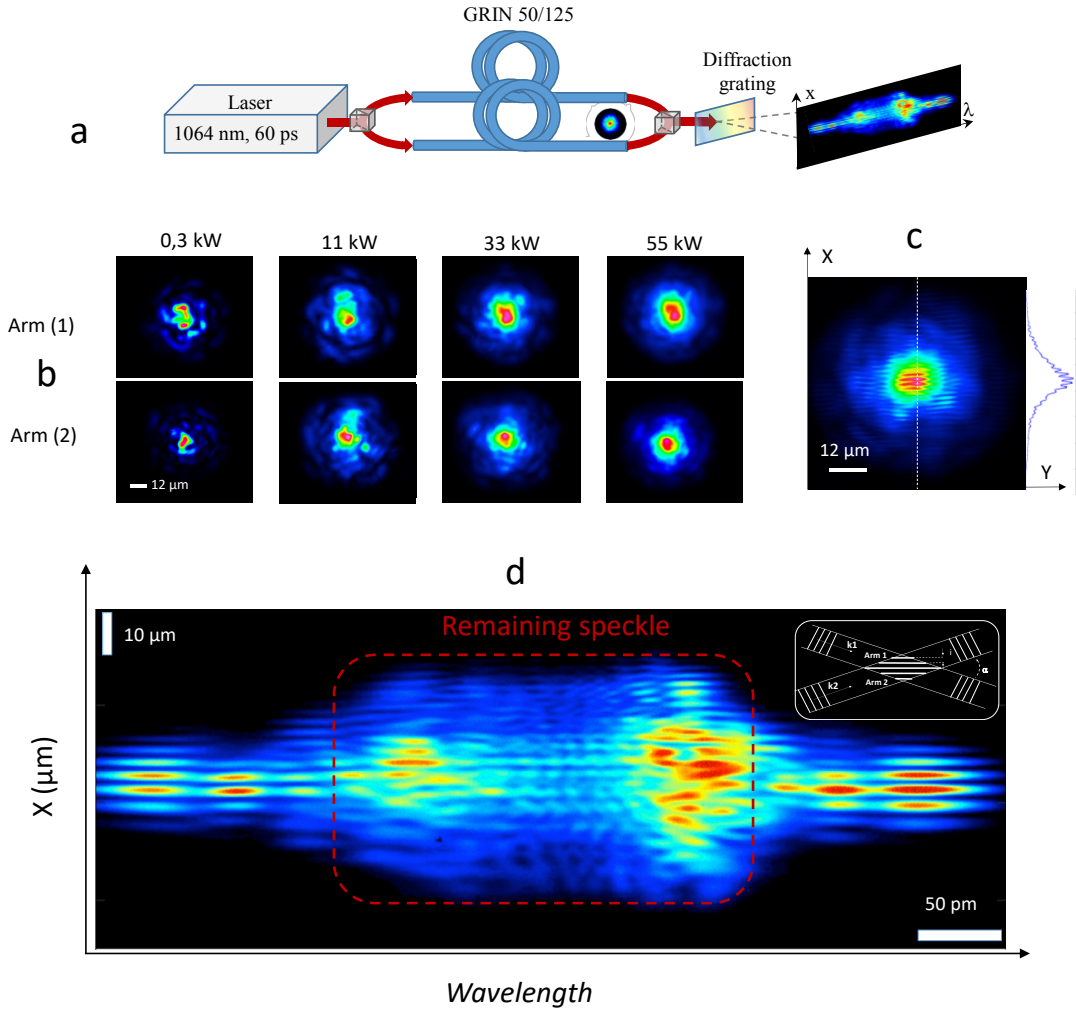

**Figure 1: Experimental evidence of phase locking between two self-cleaned beams in GRIN MMFs.** a, Schematic representation of the experimental setup as detailed in the Methods section. b, Near-field images of the output beams observed for increasing input peak power, independently recorded in the two arms of the Mach-Zehnder interferometer, seeded by the same laser source at 1064 nm. c, Spatial interference pattern produced by the two self-cleaned output beams, spatially superimposed with a 4° tilt angle (44 kW). d, Same interference pattern between the two self-cleaned output beams (obtained for 44 kW input power), but spectrally decomposed along wavelength. Note that here, the vertical axis corresponds to the X transverse dimension.

First, we investigated spatial self-cleaning that separately occurs in each of the two GRIN MMFs. In Fig.1b we display the evolution (with respect to their input power) of the 2D spatial beam profiles that we measured at the output of each fibre. At a relatively low peak power (P= 0.3 kW), an irregular transverse profile with randomly distributed speckles was visible at the output of both interferometer's arms. When we increased the input beam powers, the output transverse intensity profiles evolved into well-defined, bell-shaped structures close to the MMF fundamental mode, surrounded by low-power speckled pedestals. After recombination, the two self-cleaned beams produced clearly visible spatial fringes (Fig.1c): the local fringes' contrast in the central part of the interferometric pattern - which corresponds to the region of spatial localization of the two fundamental modes - reaches up to 22% when the self-cleaning process is well established, and it was found robust against environmental perturbations, e.g., fibre bending and squeezing.

In order to gain insight into the mutual coherence properties of the two beams, it is necessary to overcome the detrimental effect of speckles. To this end, we inserted a diffraction grating at the interferometer output, in order to separate the intense self-cleaned beam from the relatively weak multimode speckled background: here we leveraged the fact that an intense beam propagating in a GRIN MMF is subject to self- and cross-phase modulation, both of which broaden the frequency spectrum. The result is presented in Fig.1d, where the interference pattern is projected along the horizontal axis to display its wavelength structure. In the central region of the spectrum, which is circumscribed by the red dashed rectangle, the contrast of the fringes remains relatively low, locally reaching a maximum value of ~ 20%: the interference pattern here is mainly generated by the beam energy that remains within the low-power speckled multimode background. In stark contrast, the outer parts of the interferogram are mainly fed by the energy carried by the high-intensity self-cleaned portion of the beams, i.e., the fundamental modes of each fibre. As self-cleaning is based on parametric power transfer, the fundamental mode concentrates most of the output beam energy: therefore, it is predominantly subject to self-phase modulation. Indeed, the best contrast, which is directly observed in lateral portions of the interferogram, is close to 80%: this clearly demonstrates that the spatial self-cleaning phenomenon preserves well the mutual coherence of the two beams, further enabling the possibility of locking these two fundamental modes in phase. A more detailed analysis of the spatial fringe contrast and its evolution depending on power is provided in the Supplementary (Fig. SM1 and Fig. SM2).

To establish the versatility and robustness of our approach, we performed a second series of experiments for investigating the mutual coherence of the two self-cleaned beams carried by different fibre modes. Specifically, by using exactly the same optical fibres and setup, we generated two beams, self-cleaned on the LP01 and LP11 modes (see Fig.2a and Fig.2b), respectively, by only adjusting the initial beam coupling conditions into one of the fibres. In fact, the self-cleaned LP11 mode was obtained by means of slight off-axis excitation (close to 2.5°) [30]. A linear superposition of these two output beams (jointly displayed in Fig.2c) generated the well-defined interference pattern shown in Fig.2d. Similar to previous experiments, the multimode speckled background (present in the near-field pattern of both output self-cleaned beams) introduced a significant randomization in the beam superposition: we observed a strongly decreased visibility of the interference fringes in the background region, which dropped down to ~ 15 %. Remarkably however, high-contrast spatial fringes can be clearly observed on each of the two lobes of the LP11 mode intensity profile, thus attesting that a strong mutual coherence is maintained even between different self-cleaned spatial modes. More importantly, interferometric fringes observed in one of the lobes appeared to be in phase opposition with fringes obtained in the other lobe. This observation confirms the predicted antisymmetric phase structure of LP11 lobes, whereas the spatial phase of the self-cleaned LP01 exhibits homogeneous transverse profile [30]. Noteworthy, spatial self-cleaning of the LP11 mode appears to be a long-lived transient propagation state, eventually followed by back-conversion into the fundamental mode (via a dynamical attractor). This transient behaviour is experimentally and numerically demonstrated, for the first time to our knowledge, in supplementary.

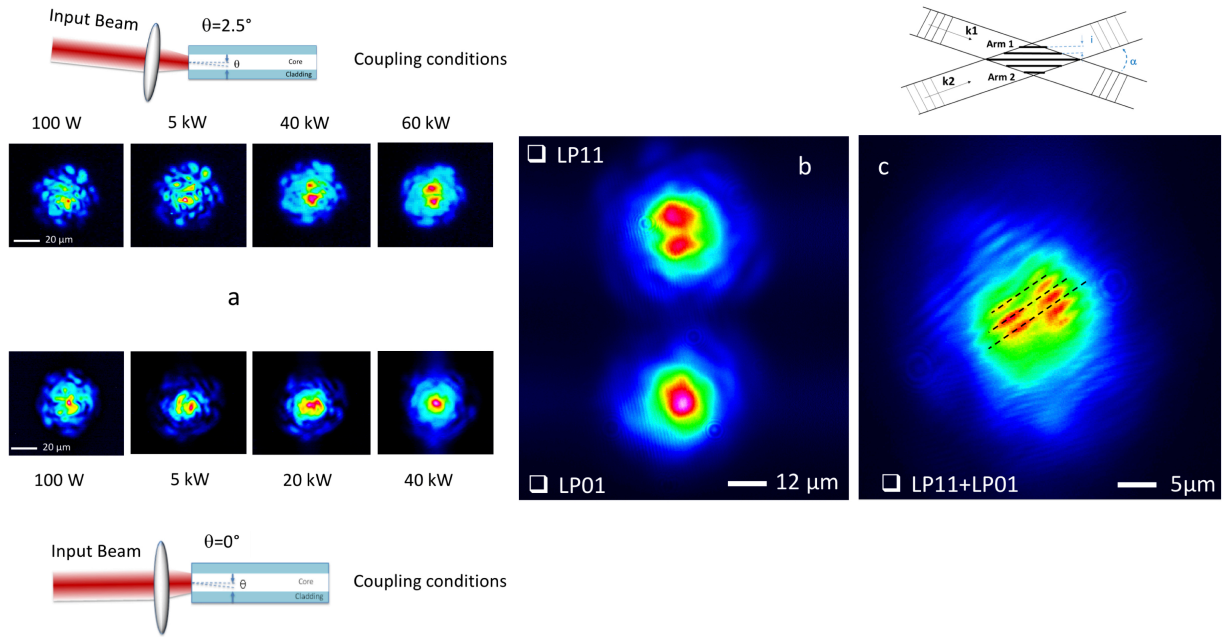

**Figure 2: Experimental demonstration of interference between two self-cleaned beams obtained in two different transverse modes.** a,b, Near-field images of the output beams for increasing input peak power, showing self-cleaning of LP11 and LP01 modes, respectively. c, Output beams self-cleaned on the LP11 transverse mode (top) and on the LP01 mode (bottom) prior to their spatial overlap. (d) Spatial interference pattern produced by two self-cleaned output beams spatially superimposed with a 4° tilt angle (dashed lines are shown as a visual guide to illustrate phase opposition between fringes from different lobes).

In order to underpin our experimental observations, we performed numerical simulations by solving a generalized version of the 3D nonlinear Schrödinger equation (GNLSE3D). Our model comprises two transverse spatial coordinates $x, y$, a time coordinate $t$, and propagation distance $z$. Fig. 3 displays the numerical evaluation of the interference pattern, which was obtained with two independent beams, self-cleaned into the LP01 modes of different fibres, with identical initial conditions. Fig.3a shows the 2D interference pattern obtained for a linear polarization state. In agreement with experiments, and as expected from spatial beam self-cleaning, the main part of the energy is localized in the central region of the interferogram. Similarly, the spectrogram of Figure 3(b) confirms that spectral broadening is mainly concentrated in the region associated with high spatial beam localization, whereas the spectral content of the speckled background remains predominantly at wavelengths around the pump.

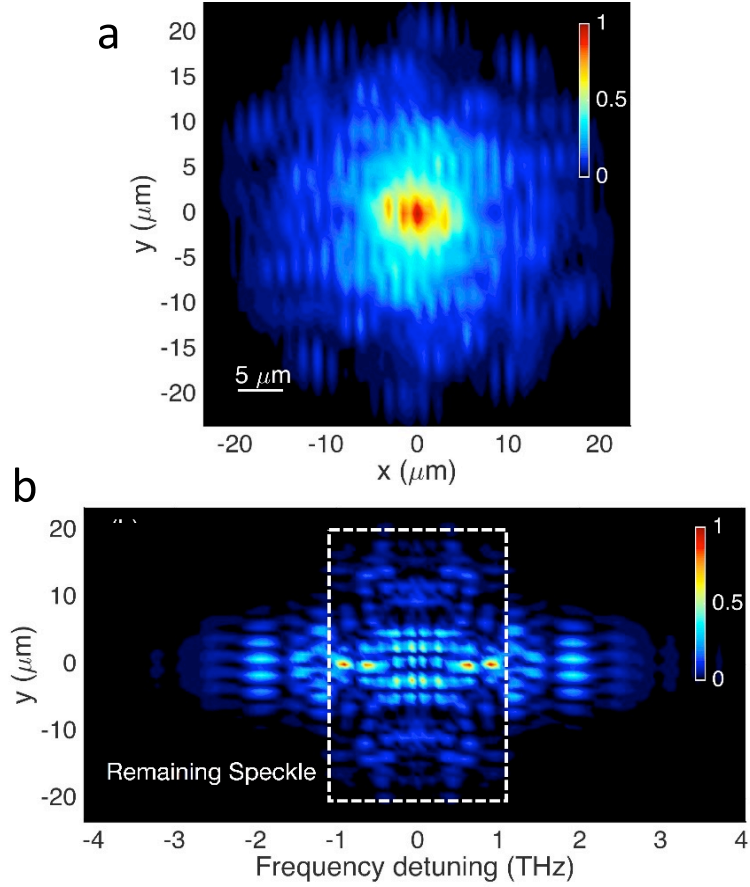

**Figure 3: Numerical simulation of interferences between self-cleaned beams obtained from two different MMFs.** a, Time-averaged 2D interference pattern, numerically generated from two independent self-cleaned beams, and considering a linear polarization filtering. b, Corresponding 1D interference pattern, projected as a function of frequency detuning from the pump.

Importantly, the interacting MMF transverse modes are initially populated by a unique Gaussian beam having a plane (transverse) spatial phase, thus equalizing all phases of the excited modes at the input-end of both fibres. In this case, quasi-monochromatic spatial parametric four-wave mixing processes (obtained in self-seeded conditions) initiate a coherent power exchange between modes, thus giving birth to an output cleaned beam with a spatial phase that can be controlled by initial conditions. As a result, a coherent and noise-free combining of the power flow carried by the two self-cleaned fundamental modes originating from two different fibres may be obtained.

In summary, we experimentally demonstrated that spatial beam self-cleaning is a purely coherent parametric process, which may generate a quasi-single-mode emission from a GRIN MMF. By providing a clear proof of the interference between two independent self-cleaned beams, we have shown that the nonlinear processes underlying spatial cleaning can preserve the mutual coherence of the two beams, even when obtained on different spatial modes. In contrast, the low-power speckled backgrounds - well visible in the output interference patterns - are characterized by complex spatio-temporal phase profiles, causing a significant decrease of the fringes' visibility upon time integration of the camera.

Besides shedding new light on fundamental aspects of beam self-cleaning and spatial condensation, the demonstrated coherence properties of spatial beam self-cleaning thus bring about interesting perspectives for coherent and scalable beam combining. Along with the potential offered by versatile mode combination, we expect that our results will pave the way toward new architectures for flexible and ultrahigh-power beam delivery.

**Methods**

**Experiments.** As a laser source, we used an amplified pulsed fibre laser, delivering 60 ps Fourier transformed pulses at 1064 nm, with the repetition rate of 300 kHz. We coupled the linearly polarized Gaussian beam into two separate spans of a GRIN MMF by using two 50 mm converging lenses and two three-axis translation stages. The beam was focused on the input face of the fibres with a diameter close to 40 μm at full width at half the maximum intensity (FWHMI). Thus, we numerically estimated that these input conditions lead to over 99% of the guided input power to be coupled into about 70 modes.
The translation stage was used to control the initial spatial coupling conditions, in order to obtain spatial self-cleaning on either the fundamental mode (LP01) or on the LP11 mode, by properly adjusting the injection angle at 0° or 2.5°, respectively. We used two 12 m-long spans of commercially available GRIN MMFs with a core diameter of 52.1μm and a numerical aperture of 0.205. The temporal broadening induced by modal dispersion (~1.5 ps for 12 m of MMF) is expected to be significantly lower than the pulse duration of 60 ps. The two MMFs were placed in the two arms of a Mach-Zehnder multimode interferometer. At the input, we used a polarizer cube followed by two half-wave plates, in order to equally split the initial beam power while adjusting the input beams polarization orientation for each MMF section. At the output, two converging lenses with a focal length of 4.5 mm and two half-wave plates were placed at the fibre ends, in order to collimate the output beams, and to adjust their polarization orientations, respectively. An output beam splitter (BS) was used to spatially superimpose the two self-cleaned beams with a small angle (~ 4°) and the resulting interference pattern was analysed with the help of a CCD camera. A delay line, implemented by means of two mirrors, was introduced in the second arm of the Mach-Zehnder interferometer, in order to synchronize the two output pulses. In addition, we used a polarizer, a bulk diffraction grating (length: 20 cm, 1800 grooves/mm) and a converging lens (f=300 mm) to spectrally resolve the interferometric pattern: the near-field profiles or the spatiotemporal patterns of the output beams were then imaged on a CCD camera through a 8 mm microlens with a magnification of 30, or through a 300 mm converging lens, respectively.

**Numerical Simulations.** We solved the GNLSE3D by considering an input Gaussian beam of 40 μm diameter FWHMI, peak intensity of 5 GW/cm$^2$ and pulse duration of 5 ps. The GNLSE includes diffraction, constant group velocity dispersion ($16.55\times10^{-27}$ s$^2$/m), waveguide contribution (radius of 25 μm, core index 1.47, cladding index 1.457) and Kerr effect ($n_2=3.2\times10^{-20}$ m^2/W and Raman fraction $f_r=0.18$). The propagation scheme was applied to the two polarization components that are incoherently coupled by the Kerr effect. Random distributed linear mode coupling was introduced with a coarse step method by applying an elliptical deformation of the fibre every 5 mm on both polarization components. The present simulations have been carried out in the absence of any Raman conversion. Further details are reported in the supplementary section.

**Acknowledgements**
M. F., M. S., A.T., and V.C. acknowledge the financial support provided by: the French ANR through the "TRAFIC project: ANR-18-CE080016-01"; the CILAS Company (ArianeGroup) through the shared X-LAS laboratory; the "Région Nouvelle Aquitaine" through the projects F2MH and Nematum; the National Research Agency under the Investments for the future program with the reference *ANR-10-LABX-0074-01 Sigma-LIM*. K. K. acknowledges the European Union's Horizon 2020 research and innovation programme under the Marie-Skłodowska-Curie (No. 713694); S. W. acknowledges the European Research Council (ERC) under the European Union's Horizon 2020 research and innovation programme (No. 740355), and the Russian Ministry of Science and Education (Grant 14.Y26.31.0017); G. M. acknowledges the Conseil Régional de Bourgogne Franche-Comté, the iXcore research foundation and the National Research Agency (ANR-15-IDEX-0003, ANR-17-EURE-0002).

**Author contributions**
M. F., M. S., Y. L., S. F., T. M., A. N. and V. C. carried out the experiments. A. T. performed the numerical simulations. All authors analysed the obtained results, and participated in the discussions and in the writing of the manuscript.

**Additional information**

**Competing financial interests**
The authors declare no competing financial interests.